\documentclass{aastex701}

\begin{document}

\title{Low-Redshift Interlopers in the SDSS DR16 $z > 5$ Quasar Catalogue}

\author[orcid=0000-0002-3277-6335,sname='Jha']{Vivek Kumar Jha }
\affiliation{National Centre for Radio Astrophysics, Tata Institute of Fundamental Research, Post Bag 3, Ganeshkhind, Pune, 411007; India}

\email[show]{vivekjha.aries@gmail.com}  

\author[orcid=0009-0009-7102-9891, sname='HK']{Harikumar N} 
\affiliation{Department of Physics and Astronomy, National Institute of Technology, Rourkela 769008, India}
\email{}

\author[orcid=0000-0002-1345-7371,sname='YW']{Yogesh Wadadekar}
\affiliation{National Centre for Radio Astrophysics, Tata Institute of Fundamental Research, Post Bag 3, Ganeshkhind, Pune, 411007; India}
\email{}

\begin{abstract}

High-redshift quasars ($z>5$) are crucial for studying early black hole growth and reionisation. SDSS Data Release 16 catalogues 655 such sources with multiple redshift estimators. We systematically examined this subsample and found 480 sources (73.2\%) where pipeline redshift (Z\_DR16Q) disagrees with the derived estimate (Z\_FIT) significantly. Visual inspection confirms that 125 of these (19.1\% of the total) are actually low-redshift active galactic nuclei. In 114 cases, broad Mg\,{\sc ii} with C\,{\sc iii} and C\,{\sc iv} was misidentified as Ly$\alpha$; in 11 cases, H$\alpha$ was misidentified as Ly$\alpha$. This error is immediately revealed by the absence of Gunn--Peterson trough blueward of Ly$\alpha$. Crucially, Z\_FIT recovers correct redshift in the affected sources where emission lines are prominent, indicating the algorithm is sound; the issue lies in community preference for Z\_DR16Q or Z\_SPEC. We urge caution against uncritical use of pipeline redshifts for $z>5$ DR16Q sources and recommend Z\_FIT as the preferred estimator.
\end{abstract}


\section{Introduction}

High-redshift quasars ($z \gtrsim 5$) represent a rare population, offering a direct observational window into the epoch of reionisation and the early assembly of supermassive black holes. Their identification, however, is subject to well-documented spectroscopic classification challenges, due primarily to the paucity of rest-frame optical features accessible within the observed optical window and the severe attenuation of flux blueward of Ly$\alpha$ by the intervening intergalactic medium \cite[IGM;][]{GP65}.

All-sky spectroscopic surveys, and in particular the Sloan Digital Sky Survey \citep[SDSS;][]{York2000}, have catalogued several hundred quasar candidates at $z > 5$. The most recent and comprehensive compilation is the sixteenth data release quasar catalogue \citep[DR16Q;][]{Lyke2020}, which provides multiple independent redshift estimates for each source. The primary catalogue redshift (\texttt{Z\_DR16Q}) is drawn preferentially from visual inspection where available, and otherwise from the automated BOSS pipeline output (\texttt{Z\_SPEC}). A supplementary catalogue by \citet{WuShen2022} provides further independent estimates for the full DR16Q sample, including an improved systemic redshift (\texttt{Z\_SYS}) and a manually verified redshift (\texttt{Z\_FIT}) using {\sc pyqsofit} \citep{Guo2018} for objects identified as exhibiting catastrophic pipeline failures.  We present here a systematic characterisation of the specific misclassification affecting the DR16Q $z > 5$ subsample, quantify its prevalence, and provide a recommendation for the appropriate redshift estimator to employ in high-redshift quasar science from this catalogue.

\begin{figure}
    \centering

    \includegraphics[width=18cm, height=9cm]{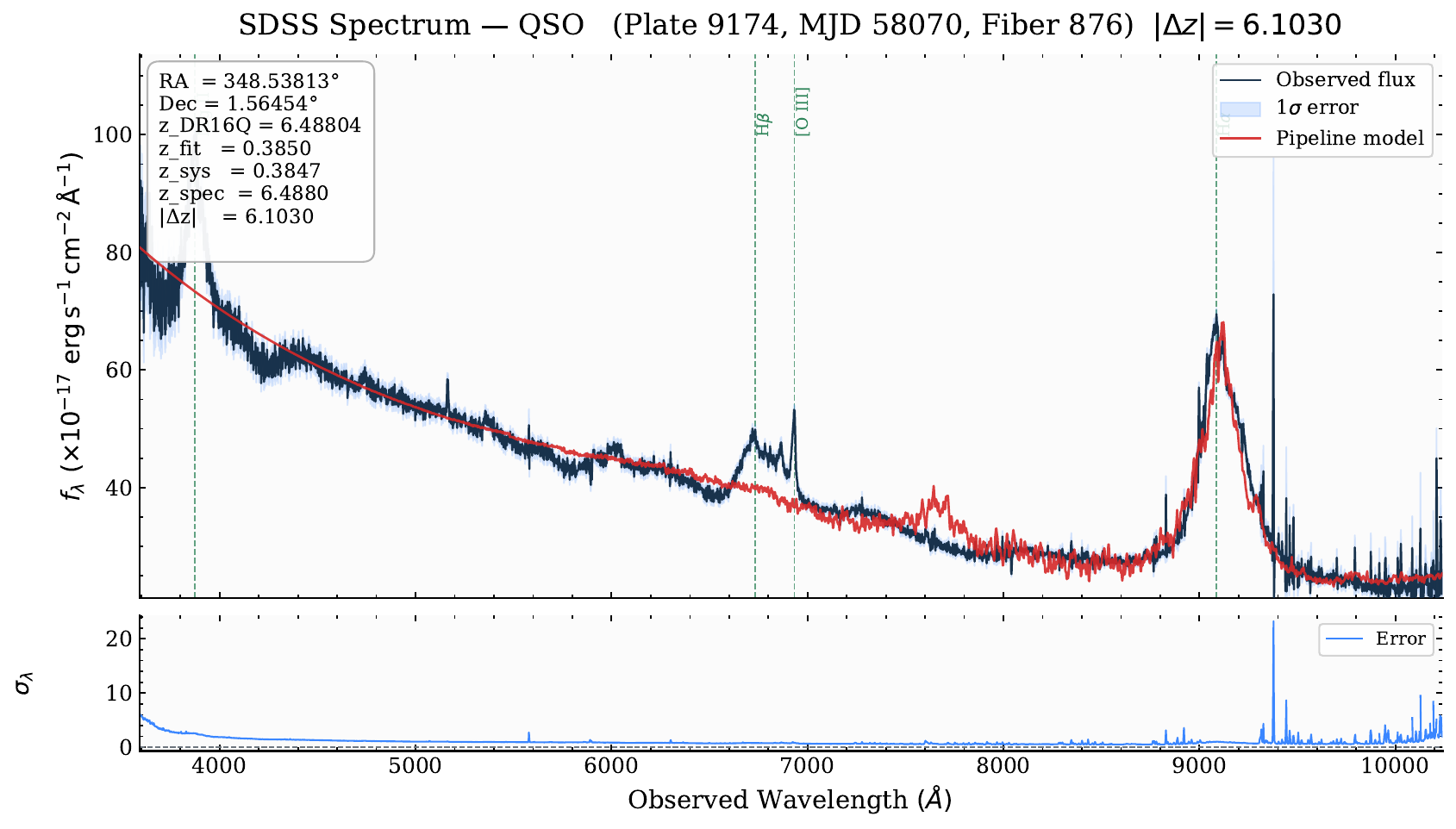}
        \includegraphics[width=10cm, height=9cm]{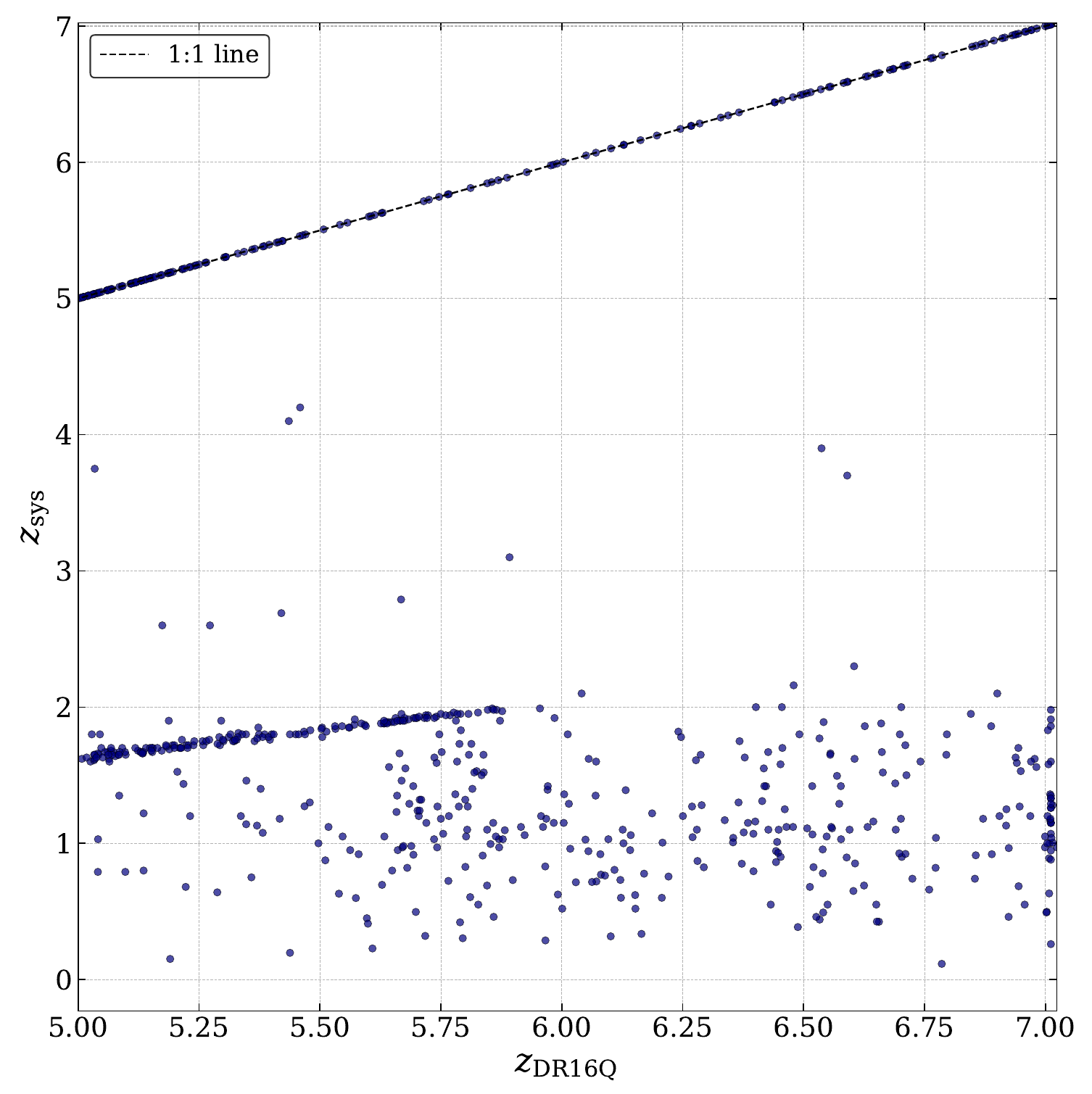}
  \caption{
\textit{Top:} SDSS spectrum of a representative 
misidentified source (Plate 9174, MJD 58070, 
Fiber 876) classified as a $z = 6.48$ quasar, but identified by the \citet{WuShen2022} 
catalogue as a low-redshift AGN at 
\texttt{Z\_FIT} = 0.385. \textit{Bottom:} \texttt{Z\_SYS} from 
\citet{WuShen2022} versus the DR16Q pipeline 
redshift \texttt{Z\_DR16Q} for all 655 
sources at $z > 5$. The dashed line denotes the 
1:1 locus.}
\label{fig:redshift_comparison}
\end{figure}

\section{Spectroscopic Misidentification in the DR16Q 
\texorpdfstring{$z > 5$}{z > 5} Subsample}

\label{sec:misidentification}

The DR16Q catalogue \citep{Lyke2020} provides 
each quasar with a hierarchy of redshift 
estimates. The primary redshift 
(\texttt{Z\_DR16Q}) is drawn from visual 
inspection where available, and otherwise from 
the automated BOSS pipeline \citep{Bolton2012}; 
for the overwhelming majority of $z > 5$ sources, 
\texttt{SOURCE\_Z\,=\,PIPE}. The supplementary 
catalogue of \citet{WuShen2022} provides an 
improved systemic redshift (\texttt{Z\_SYS}) 
from emission-line fitting, and a manually 
verified redshift (\texttt{Z\_FIT}) for objects 
with catastrophically incorrect 
\texttt{Z\_DR16Q} values.

Across the 655-source $z > 5$ subsample, 
\texttt{Z\_DR16Q} and \texttt{Z\_SPEC} exhibit 
a strict one-to-one correspondence, as do 
\texttt{Z\_FIT} and \texttt{Z\_SYS}, confirming 
two distinct estimator families: pipeline and 
fitting. Figure~\ref{fig:redshift_comparison} 
displays \texttt{Z\_SYS} against \texttt{Z\_DR16Q} 
for all 655 sources. Two populations are 
immediately apparent: a diagonal sequence of 
genuine high-redshift quasars where estimators 
agree, and a horizontal cloud concentrated at 
$z_{\rm SYS} \lesssim 2$ spanning the full 
range $5 < z_{\rm DR16Q} < 7$. Discrepancies 
of $\Delta z \sim 4$--6 within this cloud — 
far exceeding known emission-line velocity 
offsets \citep{Shen2016} — indicate categorical 
line misidentification. Following 
\citet{WuShen2022}, we adopt $\Delta z > 0.22$ 
($|\Delta V| \gtrsim 10{,}000$ km\,s$^{-1}$ 
at $z \sim 6$) to define catastrophic failure.

Visual inspection reveals two misidentification 
channels. In the first, H$\alpha$ 
($\lambda_{\rm rest} = 6563$\,\AA) is 
erroneously identified as Ly$\alpha$ 
($\lambda_{\rm rest} = 1216$\,\AA), placing 
sources at $z_{\rm true} \sim 0.33$--$0.45$ 
spuriously at $z \sim 6.2$--$6.8$. The 
misidentification is confirmed by H$\beta$ 
and [O\,\textsc{iii}] $\lambda\lambda$4959,5007 
at the correct redshift, alongside a blue 
continuum inadmissible for a genuine $z > 5$ 
quasar. In the dominant second channel, 
Mg\,\textsc{ii} ($\lambda_{\rm rest} = 
2799$\,\AA) is misidentified as Ly$\alpha$, 
placing sources at $z_{\rm true} \sim 1.5$--$1.8$ 
spuriously at $z > 5$, confirmed by 
C\,\textsc{iii}]~$\lambda$1908 and 
C\,\textsc{iv}~$\lambda$1549. In both channels, 
the misidentification is immediately apparent 
from the conspicuous absence of the 
Gunn--Peterson trough \citep{GP65} blueward 
of the putative Ly$\alpha$ wavelength.

\citet{Lyke2020} caution that $z > 5$ 
pipeline-sourced objects ``should be considered 
suspect,'' and \citet{WuShen2022} note 
qualitatively that a substantial fraction of 
$z_{\rm DR16Q} > 5$ quasars are lower-redshift 
objects. Neither publication quantifies the 
contamination rate or identifies the 
misidentification channels. Of 655 DR16Q 
sources at $z > 5$, visual classification 
confirms 125 genuine low-redshift interlopers: 
11 via H$\alpha$ and 114 via Mg\,\textsc{ii}. 
For the remainder, classification is 
inconclusive due to low S/N or ambiguous 
spectral features.

\section{Remedies and Recommendations}
\label{sec:remedies}
The misidentification propagates directly from 
the automated BOSS pipeline \citep{Bolton2012} 
into \texttt{Z\_DR16Q} \citep{Lyke2020} and 
equivalently into \texttt{Z\_SPEC}, since both 
are drawn from the same algorithmic solution. 
Any investigator employing either column 
inherits this error without indication that a 
physically consistent alternative exists within 
the catalogue infrastructure.

Of 655 DR16Q sources at $z > 5$, 480 (73.2\%) 
exhibit catastrophic redshift failures of 
$\Delta z > 0.22$, of which 125 are confirmed 
low-redshift interlopers. The correct redshift 
is already present in the catalogue via 
\texttt{Z\_FIT} \citep{WuShen2022} — a manually 
verified estimate for objects with catastrophically 
incorrect \texttt{Z\_DR16Q} values. The failure 
is therefore one of user awareness rather than 
algorithmic capability.

We advance the following recommendations. 
\textit{For users of DR16Q:} employ 
\texttt{Z\_FIT} from \citet{WuShen2022} in 
preference to \texttt{Z\_DR16Q} or 
\texttt{Z\_SPEC} for $z > 5$ science. Any 
source with $\Delta z > 0.22$ between the two 
families should be treated as a candidate 
misidentification requiring individual spectral 
verification before inclusion in any analysis.
\textit{For future catalogue releases:} we 
urge the SDSS collaboration to incorporate the 
\citet{WuShen2022} redshifts into the primary 
column for affected $z > 5$ sources, and to 
implement a Gunn--Peterson flux criterion as 
a validation step in future high-redshift 
pipeline classifications.

\begin{acknowledgments}
The authors thank Amina Thekkoth, whose enquiry about quasar redshift determination provided the impetus for this investigation. The authors thank Michael Strauss for helpful comments and discussion.

\end{acknowledgments}

\facilities{SDSS}

\software{astropy \citep{2022ApJ...935..167A} }

\bibliography{main}{}
\bibliographystyle{aasjournalv7}

\end{document}